\documentclass[electronic]{vgtc}

\graphicspath{{figures/}{pictures/}{images/}{./}}

\usepackage{times}

\usepackage{mathptmx}
\usepackage{amssymb}
\usepackage{array}
\usepackage[table]{xcolor}

\onlineid{0}

\vgtccategory{Research}

\vgtcinsertpkg

\nocopyrightspace

\newcommand\blfootnote[1]{%
  \begingroup
  \renewcommand\thefootnote{}\footnote{#1}%
  \addtocounter{footnote}{-1}%
  \endgroup
}

\title{Personalised 3D Mnemonics for Vocabulary Learning in Context}

\author{Karolina Trajkovska\thanks{e-mail: karolina.trajkovska@famnit.upr.si}\\ %
        \parbox{1.4in}{\scriptsize \centering University of Primorska} %
\and Matja{\v{z}} Kljun\thanks{e-mail: matjaz.kljun@upr.si}\\ %
     \parbox{1.4in}{\scriptsize \centering University of Primorska \\ Stellenbosch University}        
\and Maheshya Weerasinghe\thanks{e-mail: maheshya.weerasinghe@famnit.upr.si}\\ %
     \parbox{1.4in}{\scriptsize \centering University of Primorska} %
\and Klen {\v{C}}opi{\v{c}} Pucihar\thanks{e-mail: klen.copic@famnit.upr.si}\\ %
    \parbox{1.4in}{\scriptsize \centering University of Primorska \\ Stellenbosch University}}

\teaser{
  \centering
  \includegraphics[width=\linewidth]{figures/teaser_last_fixed.pdf}
  \caption{Timeline of mnemonic vocabulary learning research, spanning from foundational keyword-method studies to a proposed situated AR vocabulary learning space.}
  \label{fig:teaser}
}

\abstract{

    Augmented reality (AR) can support vocabulary learning by placing mnemonic cues in the physical contexts where new words are encountered. Prior work has augmented real objects with bilingual labels, audio, keywords, and 3D visualisations of those keywords interacting with the objects, but these cues are typically predefined by system designers. This limits scalability and reduces learners' ability to construct personal mnemonic links. By grounding mnemonics in objects that learners already recognise, use, and associate with prior experiences, a system could connect new vocabulary to existing mental models rather than requiring learners to adopt designer-defined associations. We outline a proof of concept and a storyboard of a generative AR approach in which learners create personalised 3D mnemonics by transforming familiar objects in their surroundings into situated mnemonic anchors. We position this concept within a timeline of related work and argue that recent advances in generative 3D asset creation enable a new design point: personalised 3D mnemonics that do not merely place content in the physical learning environment, but modify familiar real-world objects so that they become part of the mnemonic itself.

} 


\begin{document}

\firstsection{Introduction}

\maketitle

\blfootnote{%
  \noindent\scriptsize\itshape
  \rule{0pt}{3ex} Authors version. Contact author: karolina.trajkovska@famnit.upr.si\par
}

The keyword method is a well-established mnemonic strategy for foreign-language vocabulary learning. It supports memorisation by creating mnemonic cues: a phonological link between the unfamiliar word and a familiar-sounding keyword, and a mental imagery association of the keyword and the word's meaning~\cite{atkinson1975mnemotechnics,raugh1975mnemonic}. Although effective, the method can be cognitively demanding because learners must identify a suitable keyword, construct a meaningful association, and imagine a memorable visual~\cite{bellezza1981mnemonic,madan2014augmented}. As shown in \autoref{fig:teaser}, prior work has addressed this challenge by externalising and automating parts of the mnemonic construction process. Theories of situated cognition and motivation further suggest that learning is strengthened when tasks and new information are made personally meaningful, relevant, and connected to the learner's specific activities and contextual surroundings~\cite{pintrich2003motivational}. Augmented reality~(AR) is particularly relevant in this regard because it can place vocabulary information directly within the learner's physical environment, linking words to the objects, places, and activities in which they are encountered. VocabulARy~\cite{weerasinghe2022vocabulary} explored this opportunity by annotating real-world objects with first- and second-language labels, audio, keywords, and 3D visuals. This approach combined contextual learning with mnemonic support and showed benefits over non-AR and non-visual baselines. However, AR vocabulary systems that rely on predefined visuals remain difficult to scale. Their mnemonic materials are tied to objects and associations anticipated by system designers, limiting support for spontaneous vocabulary learning in new environments. Learners may also benefit from mnemonic associations that are personal, humorous, or culturally specific. Moreover, by grounding mnemonics in objects that learners already recognise, use, and associate with prior experiences, the system can connect new vocabulary to existing mental models rather than requiring learners to adopt designer-defined associations.

We propose extending the design space opened by VocabulARy 
with personalised 3D mnemonics. Rather than merely placing mnemonic content in the physical environment, our concept allows learners to transform familiar real-world objects so that they become part of the mnemonic itself. In our concept, learners encounter a real object in context and generate a 3D mnemonic representation in real time such that the association between the target object and the mnemonic cue is embedded in the generated asset. This concept preserves the contextual grounding of AR vocabulary learning while adding three capabilities: (i) scalability, because mnemonic objects no longer need to be created in advance; 
(ii) learner authorship, because learners can shape the mnemonic representation themselves; 
and (iii) integrated association between the keyword and the target word, because the interaction point between the real-world object and the mnemonic cue is present in the generated augmentation. Rather than only seeing a prepared mnemonic, learners can actively create one.

Building on recent advances in generative 3D asset creation, enabled by tools such as TRELLIS.2~\cite{xiang2025trellis2} and Meshy AI\footnote{https://www.meshy.ai/}, we explore AR vocabulary learning in which learners generate spatial mnemonics by augmenting elements of their physical environment. In addition to proposing this concept, we conduct exploratory experiments with a generation pipeline for producing 3D mnemonic assets from vocabulary items and contextual object cues. These experiments examine the feasibility of generating personalised mnemonic content that can be situated in AR and connected to familiar real-world objects. We position this approach within a timeline of related work and derive design recommendations for future AR vocabulary learning systems based on personalised 3D mnemonics.

This paper makes three scoped contributions: (1) a targeted synthesis of the progression from keyword-method mnemonics to personalised generative cues; (2) an exploratory, offline proof of concept showing that image-to-3D pipelines can produce candidate mnemonic assets; and (3) a conceptual interaction storyboard for an AR implementation. The paper does not present an AR system implementation. An end-to-end AR system implementation is presented separately in a companion demo-track submission~\cite{trajkovska2026vocabulary2}. Evaluation of learning outcomes remains future work.

\section{Related Work}

Interactive systems have progressively externalised and automated parts of the vocabulary learning process with the keyword method. We organise this prior work into the progression shown in~\autoref{fig:teaser}. Following the foundational keyword-method studies of the 1970s, later computational and human-computer interaction research began to automate the generation of mnemonic keywords (verbal cues) in order to make the method more accessible and scalable. Around 2014, systems such as TransPhoner~\cite{savva2014transphoner} and related NLP and computation-based approaches generated candidate keywords by considering phonetic, orthographic, semantic, and imageability-based properties of words~\cite{ ozbal2014automation, anonthanasap2015automated}. These systems shifted part of the mnemonic task from learners to computational tools: instead of inventing a cue entirely unaided, learners could choose from automatically suggested verbal cues.

The next notable step in this progression is the previously discussed VocabulARy system~\cite{weerasinghe2022vocabulary}, which shifted keyword-method support from screen-based or paper-based study materials into the learner's physical environment. Rather than introducing a new form of keyword generation, VocabulARy contributed a spatial and contextual layer: mnemonic cues were placed alongside real-world objects, connecting new words to the situations in which they are encountered. In our timeline, this work marks an important transition from automating mnemonic cues to situating those cues in AR.

Recent advances in generative artificial intelligence have moved mnemonic vocabulary learning beyond automated keyword suggestion toward the generation of complete mnemonic materials, including both verbal and visual cues. SmartPhone~\cite{lee2023smartphone} explored the use of automatically generated verbal and visual cues for keyword-method learning, marking a shift from keyword selection alone toward richer mnemonic materials. Other work has investigated AI-generated mnemonic keywords in flashcard-based learning environments~\cite{agnes2024fostering}, automatic keyword-mnemonic generation with large language models~\cite{el2024keymagine}, and overgenerate-and-rank pipelines that produce multiple candidate mnemonics~\cite{lee2024exploring}. Together, these systems demonstrate that mnemonic materials can be produced more flexibly and at larger scale than manually created ones. However, they largely remain screen-based and focus on verbal or two-dimensional visual materials. SMART~\cite{balepur2024smart} further shows how LLM-generated mnemonics can be improved through student feedback, connecting generative automation with learner-aware adaptation which presents the bridge towards the next group: Personalised Verbal and/or Visual cues.

Most recently, an emerging line of work has begun to move beyond automated generation toward personalised and human-AI co-created mnemonics. These very recent systems explore how learners can influence, adapt, or participate in the construction of mnemonic materials rather than simply receiving generated cues. Attygalle et al.~\cite{attygalle2025texttoimage} investigate text-to-image generation for vocabulary learning with the keyword method, showing how generative visuals can externalise mnemonic imagery. Lee et al.~\cite{lee2026personalizing} focus on learner preferences in adaptive Kanji mnemonics, while MnemoMaker~\cite{seow2026mnemomaker} explicitly examines different roles learners may take in human-AI mnemonic co-creation. WordCraft~\cite{shao2026wordcraft} further scaffolds the keyword method with multimodal LLMs, supporting learners across keyword selection, association building, and mnemonic material creation. Together, these works suggest a shift from automation as replacement toward AI as a collaborator in mnemonic construction.

Finally, our work identifies a new design point in this progression: \textit{Learning in AR with personalised cues}. Personalised 3D mnemonics represent one instance of this design point, where learners create mnemonic cues that are situated in their own physical environment.

\section{Generation experiments}

To validate the technical feasibility of the proposed concept, we implemented and tested two generation pipelines: experiment~1 utilises Gemini Pro for 2D image generation and modification, followed by TRELLIS.2 for 3D \texttt{.glb} object synthesis, while experiment~2 uses Meshy AI Premium for object generation. Both models (TRELLIS.2 and Meshy AI) represent state-of-the-art 3D generative models, but there are distinct trade-offs depending on deployment constraints. TRELLIS.2 is an open-source model that offers high flexibility and data privacy, but it requires significant local computational resources (e.g., a high-performance GPU) to execute efficiently. Conversely, Meshy AI provides a streamlined, cloud-based API that entirely offloads computational hardware demands, though it operates as a paid, proprietary service.

\subsection{Experiment 1: Gemini 3 Pro Image and TRELLIS.2}

\begin{figure}[h!]
    \centering
    \includegraphics[width=\linewidth]{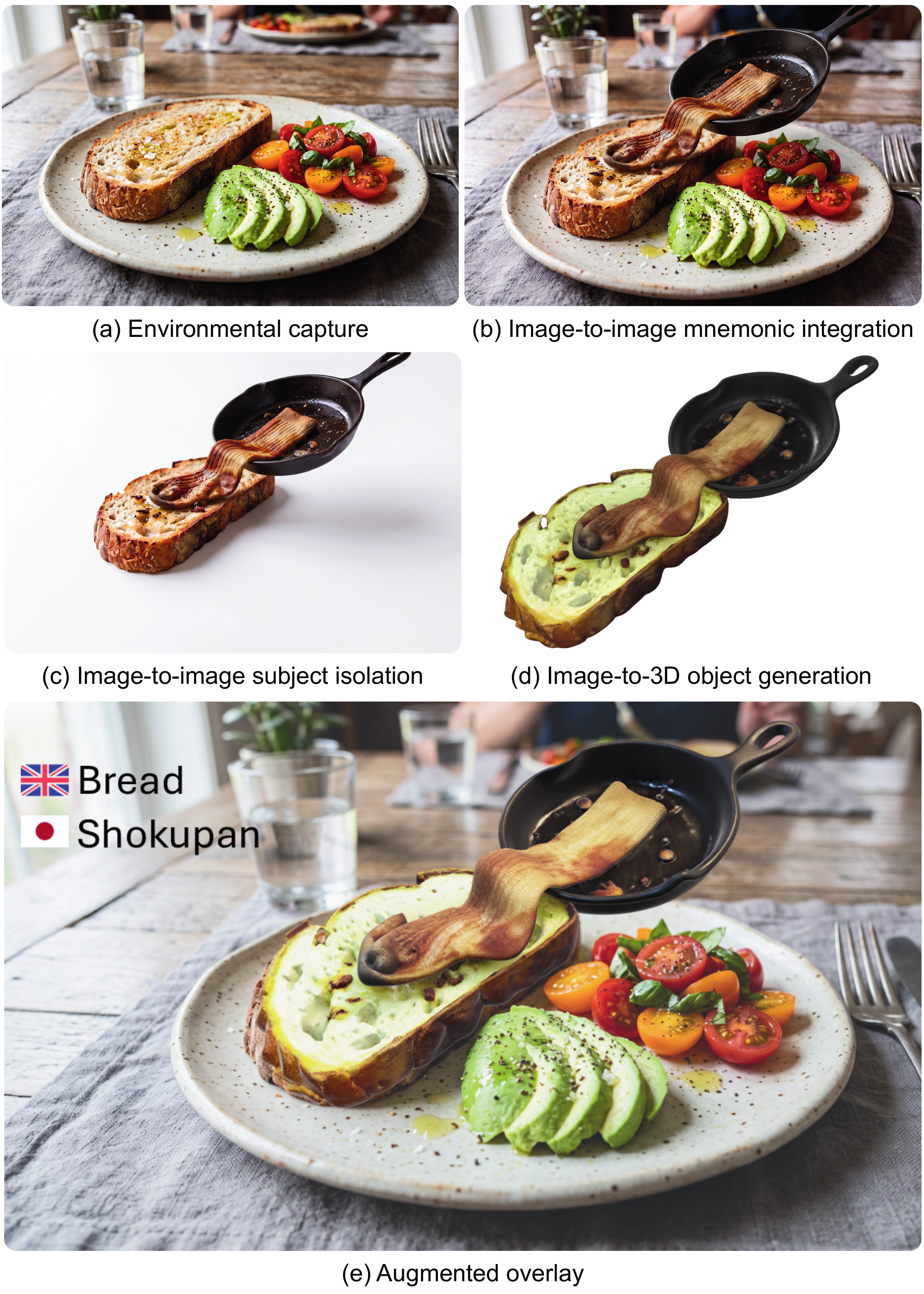}
    \caption{Stages of the generative pipeline: (a) Environmental capture. (b) Image-to-image mnemonic integration of the keywords ``sock and pan" for \textit{shokupan} (``bread" in Japanese) using Nano Banana Pro. (c) Image-to-image subject isolation. (d) 3D object synthesis using TRELLIS.2. (e) Conceptual augmented overlay of the generated \texttt{.glb} asset. \textit{NOTE: Except for (e), no manual editing was done on images.} }
    \label{fig:experiment1}
\end{figure}

The first experiment validates the local generation pipeline consisting of five steps (\autoref{fig:experiment1}):

\begin{enumerate}
    \item Environmental capture: A reference image of the learner's physical environment is captured.
    \item Image-to-image mnemonic integration: We used Gemini 3 Pro Image (Nano Banana Pro\footnote{https://docs.cloud.google.com/gemini-enterprise-agent-platform/models/gemini/3-pro-image}) to introduce a visual mnemonic into the captured scene. For this experiment, we target the English-Japanese vocabulary pair ``Bread" and ``Shokupan". To construct the mnemonic, the Japanese target word is broken down into phonetically similar English keywords: ``Sock" and ``Pan". This composite imagery is integrated into the scene with the prompt: ``On this piece of bread, generate a floating pan that is pouring a frying sock on top of the bread. The sock is sliding from the pan onto the bread." With this, the interaction point between the mnemonic cues (``Sock" and ``Pan") and the target-word (``Bread") is visible.
    \item Image-to-image subject isolation:  We used Gemini 3 Pro Image as well to isolate the mnemonic from the scene, since TRELLIS.2 performs better on isolated objects. 
    \item Image-to-3D object generation: The isolated 2D image is processed through the TRELLIS.2 image-to-3D generation pipeline. This step relies solely on the isolated image as its input (without additional text prompts) to synthesise a fully textured \texttt{.glb} asset.
    \item Augmented overlay: The resulting 3D \texttt{.glb} model is superimposed onto the original physical object within the learner's view. This was done using manual alignment of the 3D object on top of a static image (see video). Automatic alignment is feasible, but would require live spatial anchoring.
\end{enumerate}

\subsection{Experiment 2: Meshy AI}

\begin{figure}[h!]
    \centering
    \includegraphics[width=\linewidth]{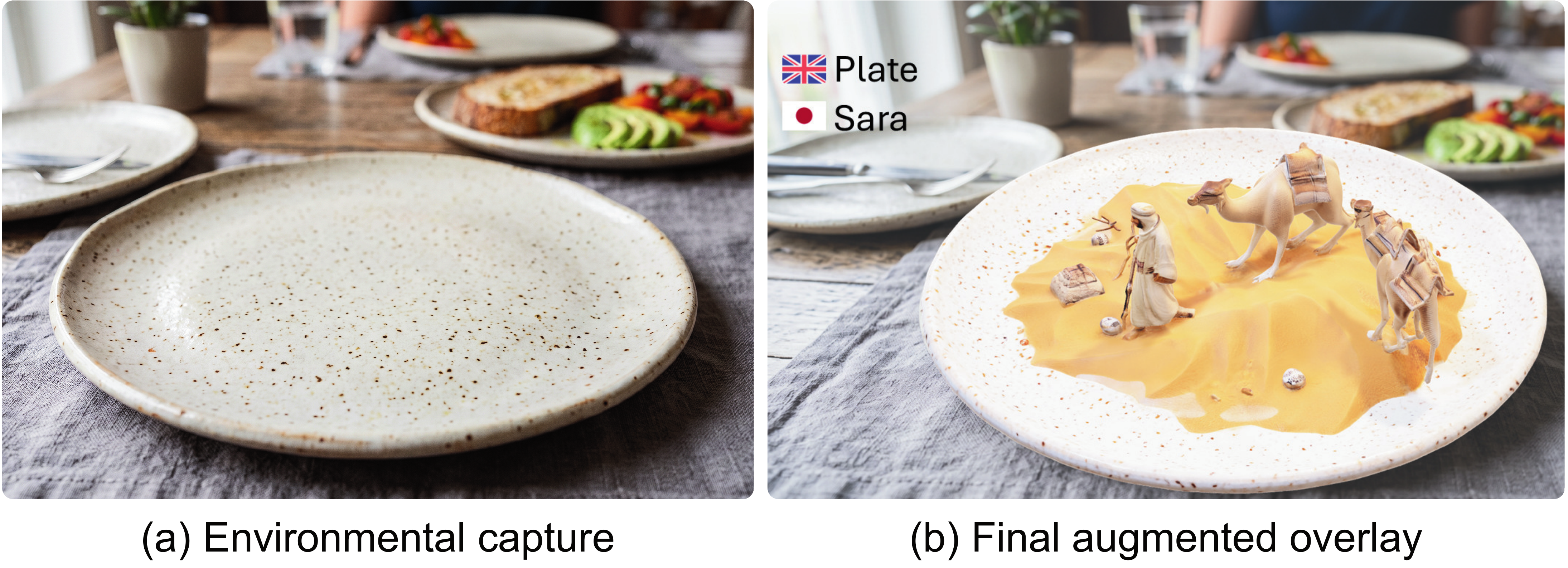}
    \caption{Results of experiment 2: (a) Initial environmental capture. (b) Final augmented overlay featuring a ``Sahara" desert mnemonic for the Japanese word \textit{sara} (plate). The 3D mnemonic was generated using the Meshy 6 model with HD Textures.}
    \label{fig:experiment2}
\end{figure}

In experiment 2, the initial 2D generation and isolation pipeline remains identical to the one used in the first experiment. However, the local 3D synthesis step is replaced with the cloud-based Meshy AI. Specifically, we used the Meshy 6 model with the HD Texture feature enabled to generate the 3D assets. The result is visible in \autoref{fig:experiment2}. This approach demonstrates the feasibility of the pipeline in hardware-constrained deployment environments where high-performance local GPU computing resources are unavailable.

\section{Interaction Design for Object-Transforming Mnemonics}

\begin{figure*}
    \centering
    \includegraphics[width=0.9\linewidth]{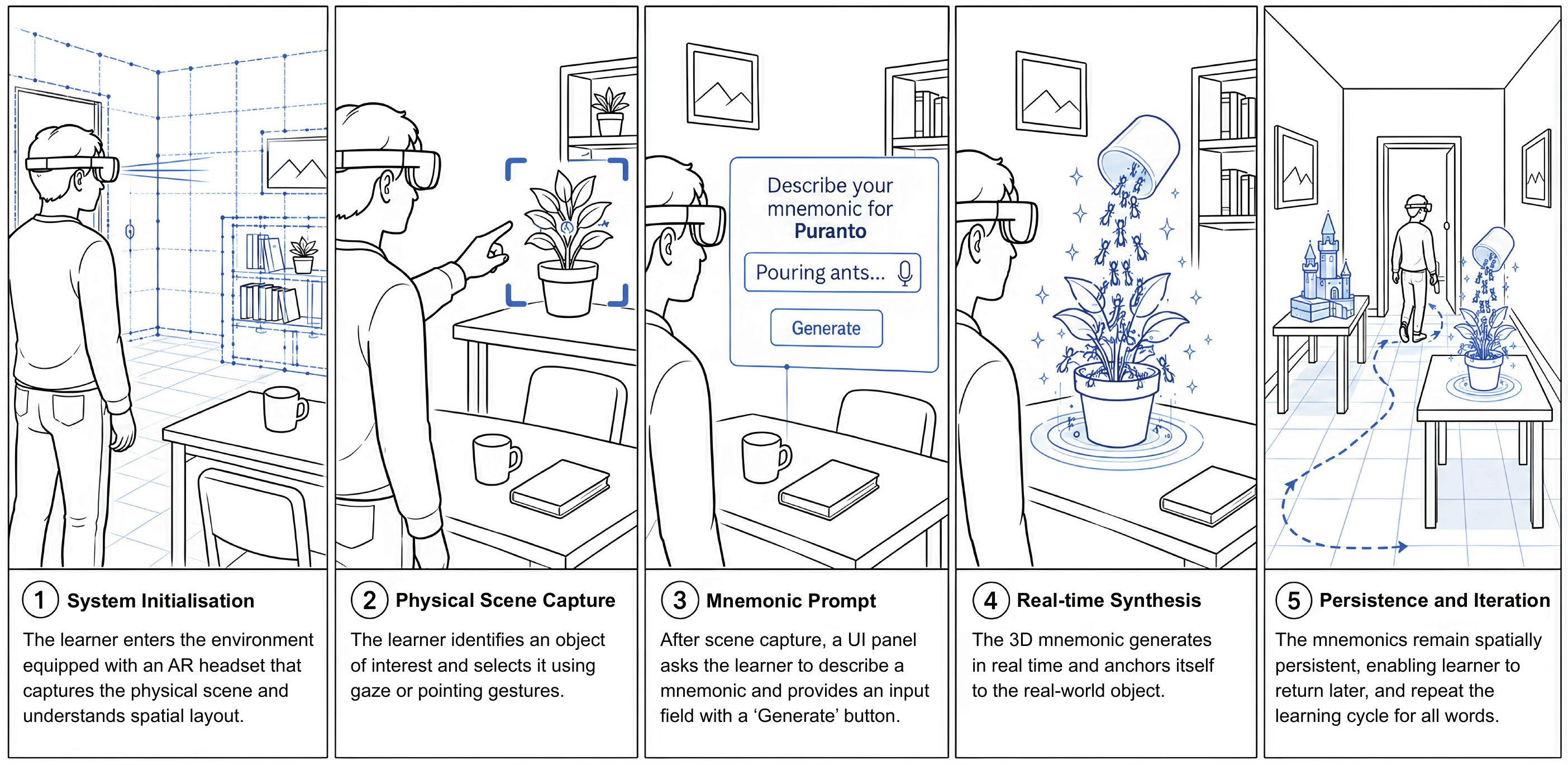}
    \caption{Conceptual storyboard illustrating the proposed interaction design for personalised 3D mnemonics in AR. The five-step pipeline includes (1) system initialisation; (2) physical scene capture; (3) mnemonic prompting; and (4) real-time 3D synthesis, resulting in (5) spatial persistence of the augmented vocabulary cue for future revision.}
    \label{fig:story}
\end{figure*}

To guide future system implementations, we propose the following conceptual interaction design illustrated in \autoref{fig:story}. The process begins with system initialisation, where the learner enters the environment equipped with an AR headset that captures the physical scene and understands the spatial layout. Next, the learner identifies an object of interest within their environment and selects it using gaze or pointing gestures. Following this, a mnemonic prompt UI panel appears, asking the learner to describe their mnemonic. This triggers the real-time synthesis stage, where the personalised 3D mnemonic is generated in real time and automatically anchored to the selected real-world object. Finally, to support persistence and iteration, the generated mnemonics remain spatially persistent, enabling the learner to return to the environment later and repeat the learning cycle for all target words.

\section{Conclusion and Future Work}

In this work, we proposed a novel generative AR concept for vocabulary learning that shifts the design space from predefined, system-authored visual cues to learner-generated, personalised 3D mnemonics. By enabling learners to transform familiar physical objects in their immediate environment into situated mnemonic anchors, our approach connects new vocabulary directly to a learner's existing mental models and contextual surroundings.
We demonstrated the preliminary technical feasibility of this concept through generation experiments using state-of-the-art models, specifically TRELLIS.2 and Meshy AI.
Finally, we presented a five-step conceptual interaction design tailored for personalised vocabulary learning in AR, outlining a comprehensive storyboard that spans from initial physical scene capture to the real-time synthesis and spatial persistence of the augmented mnemonics. A companion demo-track submission presents an example of interactive AR implementation guided by the storyboard~\cite{trajkovska2026vocabulary2}. This paper, on the other hand, focuses on the related-work synthesis, the asset-generation proof of concept, and the conceptual interaction flow.

Moving forward, future research will focus on empirical evaluations to measure the impact of personalised 3D mnemonics on both long-term vocabulary retention and cognitive load when compared to traditional methods.
To further reduce cognitive friction, the storyboard can be extended by integrating intelligent scaffolding via Large Language Models~(LLMs), proactively assisting learners in constructing highly effective and personalised mnemonic prompts as demonstrated by the related work.

\bibliographystyle{abbrv-doi}

\bibliography{template}
\end{document}